 \documentclass[final,5p,times,twocolumn]{elsarticle}

\usepackage{amssymb}
\usepackage{amsmath}
\newcommand{\ket}[1]{\mbox{$ | #1 \rangle $}}
\newcommand{\bra}[1]{\mbox{$ \langle #1 | $}}
\usepackage{graphicx}% Include figure files
\usepackage{dcolumn}% Align table columns on decimal point
\usepackage{bm}% bold math
\usepackage{natbib}

\journal{Optics Communications}

\begin{document}

\begin{frontmatter}



\title{Comment on ``A special attack on the multiparty quantum secret sharing of secure direct communication using single photons''}


%\author{}
\author[a]{Cheng-An Yen}
\author[a]{Shi-Jinn Horng}
\author[b,c]{Hsi-Sheng Goan\corref{cor1}}
\cortext[cor1]{Corresponding author.
Tel: +886 2 3366 5163;  
Fax: +886 2 2363 9984.}% Address: Department ofPhysics, National Taiwan University, Taipei 10617, Taiwan.}
\ead{goan@phys.ntu.edu.tw}
%\author{Name\corref{cor1}\fnref{label2}}
\author[d]{Tzong-Wann Kao}
%\email{D9215006@mail.ntust.edu.tw}
\address[a] {Department of Computer Science and Information
  Engineering, National Taiwan University of Science and Technology,
  Taipei, Taiwan}
%\email{goan@phys.ntu.edu.tw}
\address[b]{Department of Physics and Center for Theoretical
  Sciences, National Taiwan University, Taipei 10617, Taiwan}
\address[c]{ 
Center for Quantum Science and Engineering, National Taiwan University, Taipei 10617, Taiwan}
\address[d]{Department of Electronic Engineering, Technology and
  Science Institute of Northern Taiwan, Taipei, Taiwan}

%\address{}

\begin{abstract}
In this comment, we show that the special attack [S.-J. Qin, F. Gao, Q.-Y. Wen, F.-C. Zhu, Opt. Commun. {\bf 281} (2008) 5472.], which claims to be able to obtain all the transmitted secret message bit values of the protocol of the multiparty quantum secret sharing of secure direct communication using single photons with random phase shift operations, fails. 
Furthermore, a class of similar attacks are also shown 
to fail to extract the secrete message.

\end{abstract}


\end{frontmatter}


%\section{}
%\label{}
%\section{Introduction}

In a recent paper, Qin \textit{et al.} \cite{QGWZ} pointed out that the HLLZ protocol \cite{HLLZ} is vulnerable to a special participant attack strategy 
(hereafter called the QGWZ attack) where Alice's secret message 
can be extracted by any dishonest agent without any help of 
the other agents.
%We will call the attack presented in \cite{QGWZ} the QGWZ attack.
However, in this paper, we comment that some errors may exist 
in the QGWZ attack and thus it fails to attack HLLZ 
protocol successfully.
%\section{The HLLZ protocol}

The six main stages of the HLLZ protocol  
are {\em preparation phase}, {\em encryption phase}, {\em the first detection}, {\em encoding phase}, {\em recovery phase} and {\em the second detection}\cite{HLLZ}.
%In short, based on the linear property of the RPSO's or equivalently 
In short, based on the linear property of the quantum random phase shift operations (RPSOs) or equivalently 
random single-qubit rotations along the $y$-axis,
the encryption, encoding and recovering phase of this protocol can be expressed
as the addition/subtraction 
of the rotation angles on each photon (qubit) at different stages:
\begin{eqnarray}
\hat{U}(\theta_j)\stackrel{\ \ \hat{U}(3 m_j \pi/2)}{\Longrightarrow}
\hat{U}(\theta_j + 3 m_j\pi/2) \stackrel{\ \ \hat{U}({-\theta_j})}{\Longrightarrow} \hat{U}(3m_j \pi/2).
%\displaystyle{\sum_{\substack{n=b,c,\cdots,z}}({-\theta_n})}}{\Longrightarrow} m_j\pi.
\label{superposition}
\end{eqnarray}
Here $\hat{U}(\theta_j)$ represents a rotation on photon $j$ with 
$\theta_j=\theta_{B_j}+\theta_{C_j}+\cdots+\theta_{Z_j}$,
%an rotation angle equal to the sum of each random angle made by each agent,
and $3m_j\pi/2$ is the rotation angle made by Alice on the photon $j$
according to the bit value $m_j=0$ or $1$ of her secret message ${\bf m}$.
The measurement action in the recovery phase by the agent (say Zach), to which Alice sent the photons with encoded message, reflects 
basically whether there is a rotation of an angle $\frac{3}{2}\pi$ 
on the photon (qubit) $j$ or not.
If there is, the bit value of the message bit $j$ is $m_j=1$; 
otherwise, $m_j=0$.

%\section{The QGWZ attack}
%\label{sec:QGWZ}
Grasping the main steps of the HLLZ protocol, 
the scenario for an dishonest agent (say 
Charlie) in the QGWZ attack is to entangle the transmitted photons with prepared ancilla photons firstly,
to escape from the detection secondly, and to attain the secret message without being discovered lastly \cite{QGWZ}.
However, the errors that 
existed in the Ref.~\cite{QGWZ} will doom Charlie to fail 
to extract the secret message even if he can escape from the detection.
The first typo in Ref.~\cite{QGWZ} is in the last sentence of the first paragraph of step (E2) in the second column on page 5473. The correct sentence for that should read as ``Then Charlie sends the fake sequence $T$ to Alice.''. In other words, the sequence that was sent out by Charlie is $T$ rather than $H$ stated 
in Ref.~\cite{QGWZ}. But this typo is very minor as it is obvious from the 
subsequent description in Ref.~\cite{QGWZ} that it is the $T$ sequence that was sent out.

The possible serious error is in the expression of Eq.~(9) of Ref.~\cite{QGWZ} which was derived 
from Eq.~(7) of Ref.~\cite{QGWZ} by a controlled-controlled-($-i\sigma_y$) operation (referred to as $CY$ operation in \cite{QGWZ})
%(referred to as $CY$ operation in \cite{QGWZ})
with photons $H_j$ and $T_j$ being the control qubits and 
photon $S_j$ being the target qubit.
The action of $CY$ operation 
on computational basis states is given by 
$CY\ket{t}\ket{c1}\ket{c2}=(-i\sigma_y)^{c1\otimes c2}\ket{t}\ket{c1}\ket{c2}$, where $c1$ and $c2$ represent the control qubits and $t$ stands for the target qubit on which the $-i\sigma_y$ operation acts.  
So the correct expression for Eq.~(9) of Ref.~\cite{QGWZ} should read as
\begin{eqnarray}
\hspace*{-1.5cm}&&\ket{\Psi_3^0}_{S_jH_jTj}
=\frac{1}{2}(\cos\theta\ket{0}-\sin\theta\ket{1})_{S_j}(\ket{00}+\ket{11}-\ket{01}+\ket{10})_{H_jT_j}.
\nonumber\\
\hspace*{-1.5cm}&&		
\label{QGWZ-0-Y}
		\end{eqnarray}	
Comparing Eq.~(\ref{QGWZ-0-Y}) above with Eq.~(10) of Ref.~\cite{QGWZ}, one can find that the two $H_jT_j$ photon states in those two equations are identical rather than orthogonal.
The statement that the two states of photons $H_jT_j$ are orthogonal 
in Ref.~\cite{QGWZ} comes from comparing the
incorrect Eq.~(9) of Ref.~\cite{QGWZ} with Eq.~(10) of Ref.~\cite{QGWZ}. 
Therefore, Charlie cannot, by measuring photons $H_jT_j$, distinguish the 
state $\ket{\Psi_3^0}_{S_jH_jTj}$
of the correct Eq.~(\ref{QGWZ-0-Y}) from 
the state $\ket{\Psi_3^1}_{S_jH_jTj}$ of Eq.~(10) of Ref.~\cite{QGWZ}, 
and thus cannot extract Alice's secret bit value in practice.  
As a result, the QGWZ attack fails to extract Alice's secret message bits 
from the HLLZ protocol.

The main idea of the QGWZ attack seems (i) to use an entanglement operation to entangle each of the photon in sequence $S$ with prepared ancilla photons to escape the first detection of the HLLZ protocol, (ii) to use the inverse entanglement operation to separate the entangled states back to separable state to obtain the secret message and to escape the second detection of the HLLZ protocol. The particular entanglement operation in the Ref.~\cite{QGWZ} is $CY$ and we have shown that it fails to extract the secret message.
One may wonder whether Charlie can extract the secret with other operations instead of $CY$ in this QGWZ-like attack.
In fact, we show below that Charlie will still fail to extract the secret 
even if he can apply other entanglement operations in the QGWZ-like attacks.
Without loss of generality, suppose the photon $j$ in 
the $S$ sequence sent out by the last agent, Zach, is in the state
\begin{eqnarray}
\ket{\chi}_{S_j}=\cos\theta_j\ket{0}_{S_j}+\sin\theta_j\ket{1}_{S_j},
\label{S_initial}
\end{eqnarray}
where $\theta_j=\theta_{B_j}+\theta_{C_j}+\cdots+\theta_{Z_j}$.
%In this way, Charlie may first prepare his two-particle entangled qubits sequence in the state as 
%$\ket{\varphi}=\frac{1}{2}(\ket{00}+\ket{11}+\ket{01}-\ket{10})_{H_jT_j}$ \cite{QGWZ}, 
%and then applys a controlled-controlled-($-i\sigma_y$) operation (referred to as $CY$ operation in \cite{QGWZ})
%with photons $H_j$ and $T_j$ being the control qubits and 
%the intercepted photon $S_j$ being the target qubit. 
%After then, Charlie sends the fake photon $H_j$ to Alice in sequence.
%He can escape from the first detection by telling either a truth or a fake rotated angle according to the orthogonal measured results.
%For the sake of explicit illustration on the QGWZ attack, we can generalize Charlie's actions in the QGWZ attack in a general form,
In this way, Charlie may prepare his ancilla qubits sequence 
$A$ each in the state $\ket{\varepsilon}_{A_j}$, and apply a particular kind of 
entanglement operation $\hat{E}$ on $\ket{\chi}_{S_j}$ such that
\begin{eqnarray}
\hat{E}(\ket{\varepsilon}_{A_j} \otimes \ket{\chi}_{S_j}) = \alpha\ket{\varepsilon}_{A_j}\ket{\chi}_{S_j}+\beta\ket{\varepsilon^\bot}_{A_j}[\hat{U}(\theta')\ket{\chi}]_{S_j},
\label{E_operation}
\end{eqnarray}
where $\left|\alpha\right|^2+\left|\beta\right|^2=1$ and 
$\left\langle \varepsilon | \varepsilon^\bot \right\rangle_{A_j} = 0$,
and $U(\theta')$ is the rotation operation along the $y$ axis 
defined in (\ref{superposition}). 
After these operations, Charlie sends sequence $S$ to Alice,
and he may escape from Alice's first detection by telling the value of his 
rotation angle of either $\theta_c$ or $(\theta_c+\theta')$.
That is, when Alice selects the photon $j$ to check the security of the quantum channel used, Charlie can first measure the corresponding ancilla qubit $A_j$ in the $\{|\varepsilon\rangle , |\varepsilon^\bot\rangle \}$ basis. 
He then announces his rotation angle of either $\theta_c$ or $(\theta_c+\theta')$ when the measured result is in $\ket{\varepsilon}$ 
or in $\ket{\varepsilon^\bot}$, respectively. 
As a result, Alice will not discover Charlie's dishonest action during her first detection step.
From this point of view, the QGWZ attack of Ref.~\cite{QGWZ} is just a simplified special case of above description where Charlie's $\hat{E}$ is the controlled-controlled-($-i\sigma_y$) operation and the rotated phase shift angle $\theta'$ in 
Eq.~(\ref{E_operation}) is $-\frac{3}{2}\pi$.
In short, Charlie hence can escape from the first detection in the HLLZ protocol with this QGWZ-like attack.

After that, in the HLLZ protocol, Alice will encode her secret message bit value on each of the remaining photons in sequence $S$
with a rotation operation along the $y$-axis by an angle of $\frac{3}{2}\pi$ if the bit value is $1$. 
Otherwise, no operation is performed if the bit value is $0$.
This message-encoding operation can be expressed as
\begin{eqnarray}
\hat{U}({3}\pi/2) = -i\sigma_y=\ket{1}\bra{0} - \ket{0}\bra{1}.
\label{U_encoding}
\end{eqnarray} 
When it is applied to the state of Eq.~(\ref{E_operation}), let us consider 
the following transformation.
%For brevity, after the message-encoding operations, 
%we still use photon $j$ to express what happens next. 
%That is 
\begin{eqnarray}
\hat{U}({3}\pi/2)\ket{\chi}_{S_j}&=&\cos\theta_j\ket{1}_{S_j}-\sin\theta_j\ket{0}_{S_j} \nonumber\\
&=&\cos(\theta_j-\frac{3}{2}{\pi})\ket{0}_{S_j}+\sin(\theta_j-\frac{3}{2}{\pi})\ket{1}_{S_j} \nonumber\\
&=&\cos{\theta'_j}\ket{0}_{S_j}+\sin{\theta'_j}\ket{1}_{S_j} \nonumber\\
&\equiv&\ket{\chi^U}_{S_j},
\label{1_encoding}
\end{eqnarray} 
where $\theta'_j = \theta_j-\frac{3}{2}{\pi}$. 
Since $\hat{U}({3}\pi/2)$ of Eq.~(\ref{U_encoding}) commutes with  
$\hat{U}(\theta')$ in Eq.~(\ref{E_operation}), 
the resultant state becomes
%the two-qubit state of $S_j$ and its corresponding qubit $E_j$ in 
%the $E$ sequence, after Alice's message encoding operation, becomes 
\begin{eqnarray}
\alpha\ket{\varepsilon}_{A_j}\ket{\chi^U}_{S_j}+\beta\ket{\varepsilon^\bot}_{A_j}[\hat{U}(\theta')\ket{\chi^U}]_{S_j}
\label{1_encode}
\end{eqnarray}
%where $\ket{\chi^U} = \hat{U}(\frac{3}{2}\pi)\ket{\chi}$.
if Alice's secret message bit value is $1$.
Otherwise, if Alice's secret message bit value is $0$, 
the resultant state remains in 
the state of Eq.~(\ref{E_operation}).

After intercepting the transmitted photon again, Charlie will apply the inverse entanglement operation $\hat{E}^{-1}$ to separate the entanglement between photon $A_j$ and $S_j$ in the QGWZ-like attack.
He then will try to figure out Alice's secret message 
by making an measurement on his ancilla qubits
in the orthonormal basis of 
$\{|\varepsilon\rangle , |\varepsilon^\bot\rangle \}$. 
Since $\hat{E}^{-1}\hat{E}=I$, 
%it means that $\hat{E}^{\dagger}\hat{E}=I$.
%In other words, 
if Charlie applies the operator $\hat{E}^{-1}$ to Eq.~(\ref{E_operation}), the state of the photons $A_j$ and $S_j$ will become
\begin{eqnarray}
\hat{E}^{-1}(\alpha\ket{\varepsilon}_{A_j}\ket{\chi}_{S_j}+\beta\ket{\varepsilon^\bot}_{A_j}[\hat{U}(\theta')\ket{\chi}]_{S_j})
= \ket{\varepsilon}_{A_j} \otimes \ket{\chi}_{S_j}.
\label{0_ES}
\end{eqnarray}
It follows that 
if the operator $\hat{E}^{-1}$ is applied to Eq.~(\ref{1_encode}), 
it will become
\begin{eqnarray}
\hspace{-0.6cm}\hat{E}^{-1}(\alpha\ket{\varepsilon}_{A_j}\ket{\chi^U}_{S_j}+\beta\ket{\varepsilon^\bot}_{A_j}[\hat{U}(\theta')\ket{\chi^U}]_{S_j})
= \ket{\varepsilon}_{A_j} \otimes \ket{\chi^U}_{S_j}.
\label{1_ES}
\end{eqnarray}
Since the photon state $\ket{\varepsilon}_{A_j}$ is the same in both
Eq.~(\ref{0_ES}) and Eq.~(\ref{1_ES}), it is obvious that Charlie cannot 
distinguish whether the state of photon $S_j$ 
has been rotated by an angle of $\frac{3}{2}\pi$ or not
from the measurement result of the ancilla photons $A_j$. 
%figure out their difference from the photon $E_j$. 
%Further, because the state of photon $S_j$, i.e. $\ket{\chi}$, is unknown to Charlie, he can neither distinguish whether the state
% has been rotated $\frac{3}{2}\pi$ or not. 
%From the above description, 
Thus Charlie cannot extract the transmitted secret bits even if he can apply 
the entanglement operations defined in Eq.~(\ref{E_operation}) 
in the QGWZ-like attacks. 
%and this truth of failure in the general case should also apply to its special case.
%In other word, the QGWZ attack in Ref.~\cite{QGWZ} should be also failure to the HLLZ protocol.

In summary,
the QGWZ attack 
%which claims to be able to obtain all the transmitted secret bits of the HLLZ protocol 
as well as the QGWZ-like attacks which apply entangled operations 
other than 
the $CY$ operation fail to obtain the transmitted 
 secret message bits of the HLLZ protocol.
%of the multiparty quantum secret sharing of secure direct communication using single photons with random phase shift operations, fails. 

\section*{Acknowledgments}
H.S.G. would like to acknowledge support from the National Science
Council, Taiwan, under Grant No. 97-2112-M-002-012-MY3, 
support from the Frontier and Innovative Research Program 
%the Excellent Research Projects 
of the National Taiwan University under Grant Nos. 97R0066-65 and 
 97R0066-67,
and support from the focus group
program of the National Center for Theoretical Sciences, Taiwan.
C.A.Y. and S.J.H. would like to acknowledge support from the National Science Council, Taiwan,
under Grants Nos. 97-2221-E-239-022- and 95-2221-E-011-032-MY3.
%\end{acknowledgments}


\end{document}